# Orientational order in liquid crystal surface induced by evaporation of water droplet

*AUTHOR NAMES*

*Yusaku Abe and Yu Matsuda*

AUTHOR ADDRESS

Department of Modern Mechanical Engineering, Waseda University, 3-4-1 Ookubo, Shinjuku-ku, Tokyo, 169-8555, Japan.

ABSTRACT

 The evaporation of a droplet induces variety of ordered patterns near the contact line between the droplet and a substrate. This pattern formation involves both the behavior of colloidal suspensions and interactions between a droplet and a substrate. Although studies on the effect of the behavior of colloidal suspensions on the deposition process have been actively conducted, studies on the effect of the interaction between a droplet and a substrate have not been actively conducted because the observation of the interaction is difficult owing to the lack of structural changes in the substrate by the interaction. In this study, we investigated the solvent-substrate interactions by using a water droplet and liquid crystals as a substrate, where the liquid crystals were placed on a glass slide. The orientation patterns of the liquid crystals were induced at the contact line between the water droplet and the liquid crystal and were observed using a polarized optical microscope. Pinning and depinning events were observed at a certain anchoring strength of the liquid crystal on the glass slide and formed linear patterns on the surface of the liquid crystal. Additionally, the interfacial tension between the water and liquid crystal slowly varied at this anchoring strength owing to the variation in the orientation of the liquid crystals. These results indicated that the evaporation-induced patterns appeared when the strength of the solvent-substrate interactions was neutral. The interaction between a droplet and a substrate is important for the fabrication of desired ordered patterns.

# INTRODUCTION

Pattern formation induced by the evaporation of droplets has been used to assemble the microstructures.[1] When droplets containing colloidal suspensions evaporate, these suspensions collect near the contact line owing to convection and capillary flow.[2–6] This local increase of the density of the suspensions will induce the pinning of the contact line and deposition of suspensions.[4] Coffee ring effect is a well-known example of this phenomenon, in which colloidal particles accumulate near the contact line due to capillary flow and deposit in circular pattern along the contact line.[1,7,8]

Understanding and controlling pattern formation by the evaporation of droplets is of interest because patterning by the evaporation can easily form ordered structures.[9–11] In particular, patterns formed during the evaporation are diverse for anisotropic colloidal suspensions.[12–15] For highly anisotropic colloidal suspensions, a local increase in the concentration of the suspensions near the contact line will induce phase transition from isotropic to nematic phase. For example, in the evaporation of droplets containing amyloid fibril rigid filamentous colloids, these colloids may deposit in the nematic phase near the contact line.[12] The evaporation of droplets containing highly anisotropic materials, such as carbon nanotubes,[16,17] DNA,[18,19] and porphyrin trimers,[20] forms various ordered patterns. It is considered that these patterns mainly depend on the behavior of colloidal suspensions and solvent-substrate interactions. The behavior of colloidal suspensions has been studied extensively. The concentration, particle aspect ratio, and evaporation rate have been known to affect the pattern formed.[8,21] However, it is difficult to evaluate the effect of the solvent-substrate interactions on the pattern formation because the observations of the effects are difficult.

This difficulty depends on the lack of structural changes in the general solid substrate, such as glass slides and mica, during evaporation.

In this study, we investigated the solvent-substrate interactions during the evaporation of diluted water droplet on a substrate of liquid crystals, which was placed on a glass slide. Observing the evaporation of the water droplet on liquid crystal, we studied the structural changes in the substrates (liquid crystals) using polarized optical microscopy (POM). Additionally, strength of interfacial interactions between solvents and substrates at the contact line can be varied by changing the anchoring strength of the liquid crystal molecules on the glass slide. We successfully observed the changes in the solvent-substrate interactions as the pinning strength of the contact line between the droplet and the liquid crystal substrate. We investigated the pinning and depinning events at the contact line in various anchoring strength of the liquid crystal. We found that the solvent-substrate interactions changed the behavior of the contact line during the evaporation. Under certain conditions, patterns were formed on the surface of liquid crystals by changing the orientation of the liquid crystal molecules on the surface. These results suggest that evaporation forms patterns solely by solvent-substrate interactions, even in the absence of colloidal suspensions inside the droplets.

**EXPERIMENTAL SECTION**

**Materials**

The liquid crystal 4-Cyano-4'-pentylbiphenyl (5CB, 98% pure) and trichlorooctadecylsilane (85% pure) was purchased from Tokyo Chemical Industry Co., Ltd, Japan.

Isopropyl alcohol, (IPA, 99.7% pure), methanol (99.8% pure), chloroform (99.0% pure), and Contaminon® were purchased from Fujifilm Wako Pure Chemical Corporation, Japan. Purchased chemicals were used as received without further modification or purification. Water used in all experiments was purified using a Milli-Q water purification system (Direct-Q UV3; Merck, Germany). Mirco slide glasses (26 mm by 76 mm by 0.8-1.0 mm) were purchased from Matsunami Glass Ind., Ltd.

**Preparation of hydrophilic and hydrophobic glass slides**

We prepared hydrohalic and hydrophobic glass slides to control the anchoring strength of the liquid crystals on the glass slide by varying the contact angle between the glass slide and the liquid crystals. The surface modification of glass slide was conducted by following procedures. First, glass slides were rinsed with IPA and water. Then, the cleaned glass slides were washed with Contaminon® using a bath-type sonicator (Bransonic® M1800Yamato Scientific Co., Ltd., Japan) to remove organic matter. Subsequently, the obtained hydrophilic glass slides were dried and stored in methanol. Hydrophobic glass slides were prepared from these hydrophilic glass slides. Hydrophilic glass slides were rinsed with chloroform to remove methanol. Then, these glass slides were surface modified with silane-coupling in chloroform; trichlorooctadecylsilane was used as a silane coupling agent.

**Observation of evaporation behavior of water droplets**

The 5CB droplet placed on a hydrophilic/hydrophobic glass slide was used as a substrate. We kept our experimental system at 20°C; the 5CB droplets exhibit a nematic phase in this condition. The contact angle of a static 5CB droplet of 1.5 μL was measured by the sessile drop method using a commercial contact angle meter (Dropmaster DMs-401; Kyowa Interface Science

Co., Ltd., Japan) to confirm the anchoring strength. Then, a water droplet of 0.5 µL was placed on the 5CB droplet using the syringe of the contact angle meter.

We imaged the evaporation behavior of water droplet on 5CB droplet surface using a polarized light microscope (BX53M; Olympus Co. Ltd., Japan) with an objective lens of 10×, N.A. of 0.25, and W.D. of 21 mm (LMPLFLN10X, Olympus Co. Ltd., Japan). We can visualize whether liquid crystal has orientational order or not by using both a polarizer and an analyzer. We can also investigate the orientation of liquid crystal molecules because more polarized light is absorbed when molecular stacking is perpendicular to the light polarization direction. Cross-polarized and λ plate images and movies were taken with a CMOS camera (Floyd multi interface digital camera; WRAYMER inc., Japan). The movies with frame rate 28.6 Hz were recorded with 35 ms exposure time for 30 min.

## RESULTS AND DISUCSSION

### Relationship between contact angle and anchoring strength of 5CB

In our experiments, we used 5CB exhibiting a nematic phase at 20°C. Figure 1a-c shows the shape dependence of 5CB droplets on glass slides with different wettability. The contact angles of a 5CB droplet were 44° (hydrophobic glass slides), 20° (untreated glass slides), and 10° (hydrophilic glass slides), respectively. By considering the balance of forces at the contact line, the interfacial tension at interface between glass slides and liquid crystal $\gamma_{\mathrm{GL}}$ is given by

$$\gamma_{\mathrm{GL}} = \gamma_{\mathrm{G}} - \gamma_{\mathrm{L}} \cdot \cos\theta \qquad (1)$$

where $\theta$ is the contact angle between the 5CB droplet and the glass slide, and $\gamma_G$ and $\gamma_L$ are surface tension of glass slides and liquid crystal, respectively. In this study, $\gamma_G$ and $\gamma_L$ are constant at each wettability condition because these properties are substance-specific values. As the wettability increases, surface tension $\gamma_{GL}$ decreases. The anchoring strength of 5CB on the glass slide decreases as a decrease in the surface tension at interface between glass slides and liquid crystal $\gamma_{GL}$. Then, the anchoring strength of 5CB decreases as an increase in the wettability. Figure 1d is a typical POM image around the edge of 5CB droplet on a glass slide when the contact angle of 5CB droplets was 44°. The birefringence on a glass slide was observed, indicating that the 5CB molecules at the surface aligned. Figure 1e and f are a bright-field microscopy image and a POM image of a contact line between diluted water and 5CB, respectively. In the bright-field microscopy image, color difference between water and 5CB is negligible. On the other hand, in the POM image, color is different between water and 5CB because water molecules were not aligned but 5CB molecules at the surface were aligned.

**Characterization of evaporation of the water droplets**

We performed POM visualization during the evaporation of a diluted water droplet on a 5CB substrate with increasing the contact angle of 5CB to glass slides, specifically 44°, 20°, and 10°. As the contact angle decreases, the anchoring strength of 5CB decreases. Then, as the anchoring strength decreases, interactions between a water droplet and a liquid crystal substrate becomes stronger. As illustrated in Figure 2, water droplets evaporated over time, while 5CB droplets didn't evaporate because of their non-volatility.

For the lowest wettability of 44° (Figure 2a-d), the radius of the water droplet decreased with time. The evaporation of the water droplet did not induce the orientation change of 5CB molecular as the color of a 5CB droplet in the POM image did not change. This result suggests that solvent-substrate interactions are weak. For the contact angle of 5CB to a glass slide of 20°, we observed changes in both the water droplet radius and orientation structure of the 5CB molecules (Figure2e-h). As seen in Figure 2e and f, while the water droplet radius hardly changed in the first 10 min, the radius rapidly decreased at 20 min. We also observed the change of the orientation structure of the 5CB molecules, which recognized as patterns in a POM image. The color of the POM image was different from the surrounding area at the position where the initial contact line existed after 20 min (Figure 2g). Moreover, three radial lines appeared as the contact line receded. These results indicate that the orientation of the 5CB molecule is affected by the evaporation of the water droplets when anchoring strength of 5CB decreased due to the higher wettability with the contact angle of 20°. It is considered that the pattern at the initial contact line position was formed by the strong solvent-substrate interactions caused by the pinning of the contact line. The radial lines appear to be line defects generated by the 5CB defects near the initial contact line and pulled during shearing these defects in the evaporation after the depinning of the contact line. For the highest wettability with the contact angle of 10° (Figure 2i-l), the contact line was not moved during the experiment (30 min). This is because the pinning strength was increased due to the stronger solvent-substrate interactions.

**Movement of the contact line**

Upon conducting experiments under various wettability conditions, we were able to demonstrate that the movement speed of the contact line changed depending on the wettability of the glass slides. Figure 3 shows the time-series variation in the normalized radius of the water droplets at different wettability conditions. When the contact angle of 5CB to glass slides was 44°, the lowest wettability, the radius of the water droplet monotonically decreased. On the other hand, when the contact angle of 5CB to glass slides was 10°, the highest wettability, the radius of the water droplet was almost constant. When the contact angle of 5CB to glass slides was 20°, medium wettability, radius of the water droplet was almost constant for the first 15 min, and then decreased monotonically.

These results show that the pinning strength between the water droplet and the 5CB droplet depends on the wettability of the 5CB to the glass slide. That is, the pinning strength increases as the wettability of glass slides increases and the anchoring strength of 5CB decreases. In conjunction with the POM images (Figure 2), pinning and depinning occurred continuously alternatively when the solvent-substrate interaction was neither too strong nor too weak; the contact angle of 20° in this study. We also found that repeat of pinning and depinning may induced two types of patterns on the surface of liquid crystal: one is the pattern along the contact line, and the other is the radial pattern perpendicular to the contact line (Figure 2, Figure 3).

**Orientation of the liquid crystal molecules**

When solvent-substrate interaction satisfies certain strength conditions, patterns are induced on the substrate surface owing to the solvent evaporation. In this study, this condition was satisfied when the contact angle of 5CB to glass slides was 20°. Figure 4 shows the polarized light transmission images of induced patterns with the $\lambda$-plate. Figure 4a and b are the images of the same location with polarization planes perpendicular to each other. In Figure 4, two types of patterns were recognized: one is the pattern along the initial contact line position, and the other is the radial pattern perpendicular to the initial contact line. The high transmittance on the pattern along the initial contact line is observed in Figure 4a, indicating that 5CB molecules were oriented parallel to the initial contact line. On the other hand, the high transmittance on the radial pattern along the evaporation direction is observed in Figure 4b, indicating that 5CB molecules were oriented parallel to the evaporation direction. These results indicate that liquid crystal molecules are oriented in a particular direction in the pattern formation.

**Configuration of the contact line during evaporation**

Pattern formation induced by evaporation is affected by solvent-substrate interactions. In this study, considering the force between water and liquid crystal phases at the contact line, we investigated the solvent-substrate interactions. The contact angles $\alpha$ and $\beta$ are defined as illustrated in Figure 5a and b, where $\gamma_{w}$, $\gamma_{LC}$, and $\gamma_{w-LC}$ are the surface tension between water and surrounding air, that between liquid crystal and air, and the interfacial tension between water and liquid crystal, respectively. It is noted that the shape of the water droplet is a stacked form of a spherical cap. The following equation is obtained from the horizonal force balance

$$\gamma_{LC} \cos\alpha = (\gamma_{w} + \gamma_{w-LC}) \cos\beta \,. \qquad (2)$$

Assuming that the variations of $\gamma_{LC}$ and $\gamma_{w}$ were small compared with those of $\cos\alpha$ and $\cos\beta$ in our experiments, we obtain the following relation:

$$\gamma_{w-LC} \propto \frac{\cos\beta}{\cos\alpha}. \qquad (3)$$

Hence, considering the ratio of $\cos\beta$ to $\cos\alpha$, we can evaluate the strength of surface tension between the water droplet and the 5CB substrate. Figure 5c shows the time-series variation of the ratio $\cos\beta / \cos\alpha$ at different wettability of the 5CB to the glass slide (44°, 20°, and 10°). Comparing the time histories of the ratios with those of the droplet radius shown in Figure 3, we discuss the dependence of $\gamma_{w-LC}$ on the motion of the contact line and the wettability of the 5CB to the glass slide. The variation in the ratio for the contact angle of the 5CB to glass slides of 44°
was greater than that of 10°. For the contact angle of 5CB to glass slides of 20°, the variation in
the ratio slowly decreased before the contact line moved (first 15 min). Subsequently, the ratio
sharply decreased after the contact line moved (after 15 min). These results indicated that the
surface tension was unstable for high anchoring strength (the contact angle of the 5CB to the glass
slide was large), and vice versa. We also found that the interfacial tension $\gamma_{w-LC}$ was stable when
the contact line was strongly pinned for the contact angle of the 5CB to the glass slide of 10° and
20° (only for first 15 min). In cases where movement of the contact line and pinning occurred
alternately, the interfacial tension $\gamma_{w-LC}$ gradually decreased and the pinning was released when
$\gamma_{w-LC}$ fell below a certain threshold. We revealed that the movement of the contact line could be
evaluated by the surface tension between the water droplet and the 5CB substrate.

Figure 5c shows the time-series of the normalized volume of the water droplets at different
wettability of the 5CB to the glass slide (44°, 20°, and 10°). The rate of decrease of the normalized
volume was greater when the contact line moved continuously. This rate for the contact angle of
20° drastically decreased during the movement of the contact line. These results suggest that the
evaporation rate is faster when the contact line is not pinned than when the line is pinned.

**CONCLUSIONS**

In summary, we investigated pattern formation during solvent evaporation by focusing on solvent-
substrate interactions. The water droplet was placed on the liquid crystals of 5CB, which was
placed on a glass slide. The patterns of the 5CB orientation induced by the evaporation of the water
droplet were visualized by a POM. Experiments were performed at various interfacial tensions between water and 5CB by changing the anchoring strength of the liquid crystals on the glass slide. At a certain surface condition, pinning and depinning of the contact line between the water droplet and the 5CB substrate alternately occurred, and the water droplet induced highly ordered linear patterns on the surface of the 5CB. In these linear patterns, the 5CB molecules oriented in a specific direction; Patterns also had highly order at the molecular scale. We found that not only the properties of colloid suspensions but also pinning and depinning events determined by solvent-substrate interactions have a strong influence on the pattern formation induced by the evaporation of solvent. The findings of this study may contribute to the further control of self-assembly by solvent evaporation and the development of highly ordered and functional materials.

## AUTHOR INFORMATION

### Author Contributions

The manuscript was written through contributions of all authors. All authors have given approval to the final version of the manuscript.

Yusaku Abe: conceptualization, data curation, formal analysis, funding acquisition, investigation, methodology, visualization, writing-original draft, and writing-review & editing. Yu Matsuda: investigation, methodology, supervision, writing-original draft, and writing-review & editing.

### Funding Sources

This work was partially supported by JST, ACT-X Grant Number JPMJAX21A, Japan.

## ACKNOWLEDGMENT

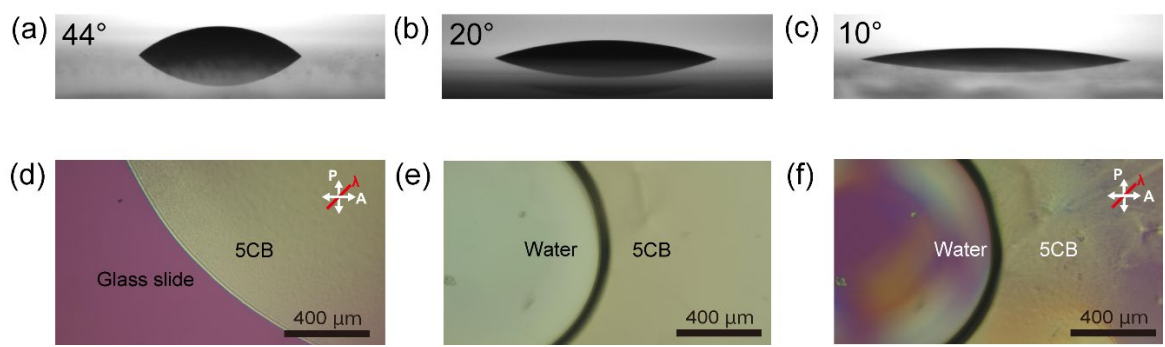

Figure 1. (a-c) measurements of contact angle of 5CB droplets on different wettability glass slides, 44° (a), 20° (b), and 10° (c), respectively. (d) a POM image around the edge of 5CB droplet on a glass slide. (e, f) a bright-field microscopy image and a POM image of a contact line between water and 5CB.

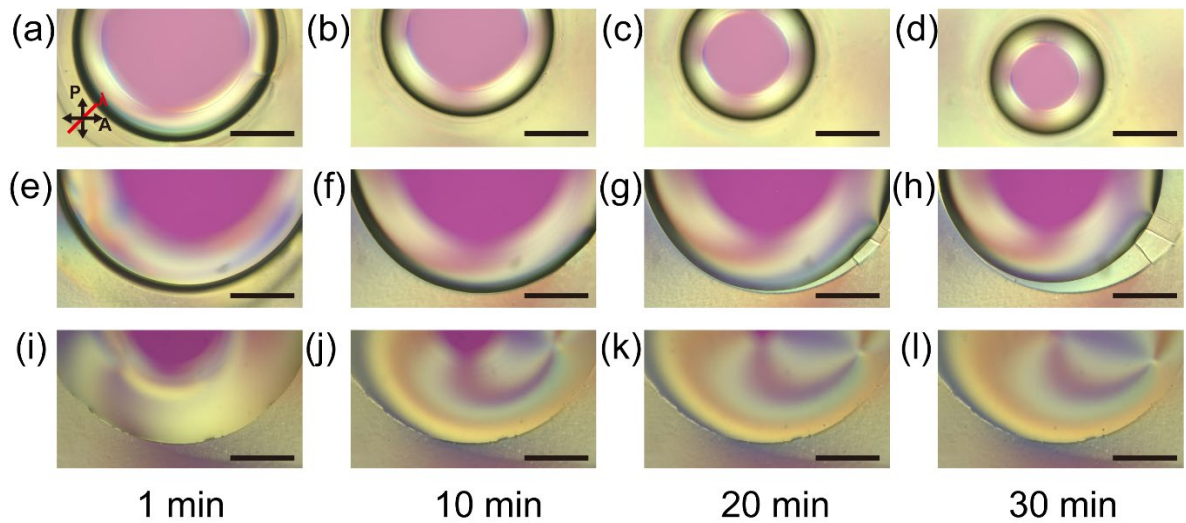

Figure 2. POM images every 10 min after dropping water at different wettability, 44° (a-d), 20° (e-h), and 10° (i-l), respectively. Scale bars are 200 μm.

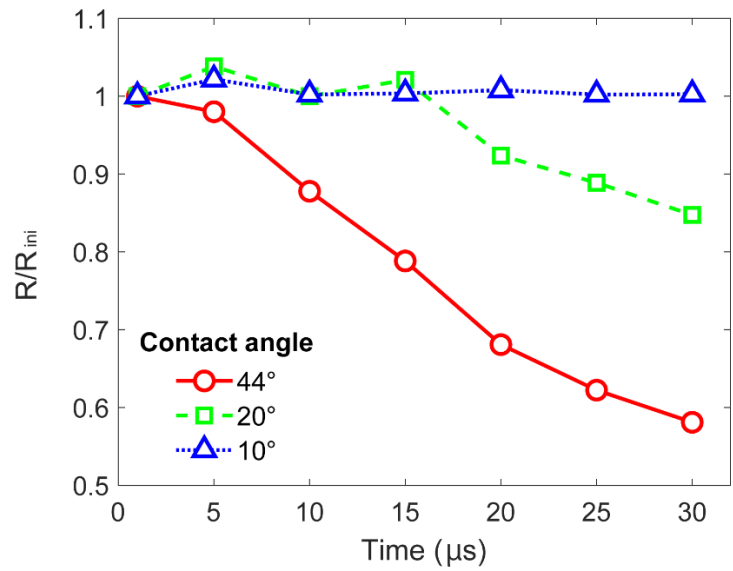

Figure 3. Normalized radius of a water droplet as a function of time for each wettability of glass slides. $R$ and $R_{ini}$ are radius of a water droplet and radius of a water droplet at 1 min, respectively.

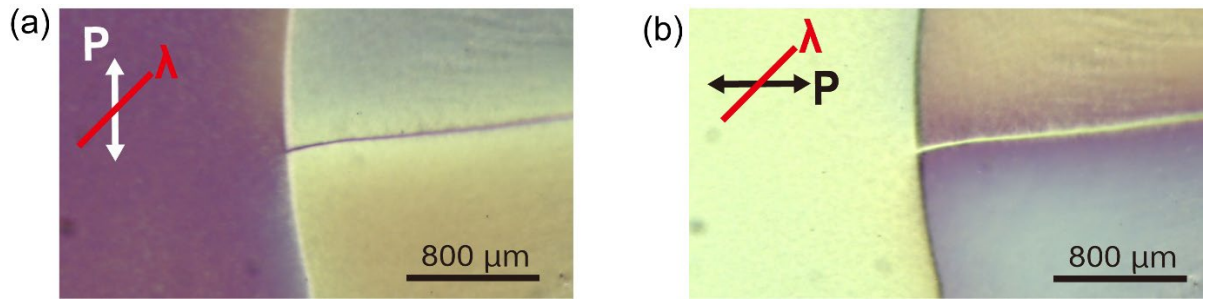

Figure 4. Polarized light transmission images around the contact line with $\lambda$-plate. (a) Direction of polarized light is vertical. An initial contact line is identified by black arrows. (b) Direction of polarized light is horizontal. Direction of movement of the contact line is identified by a white arrow.

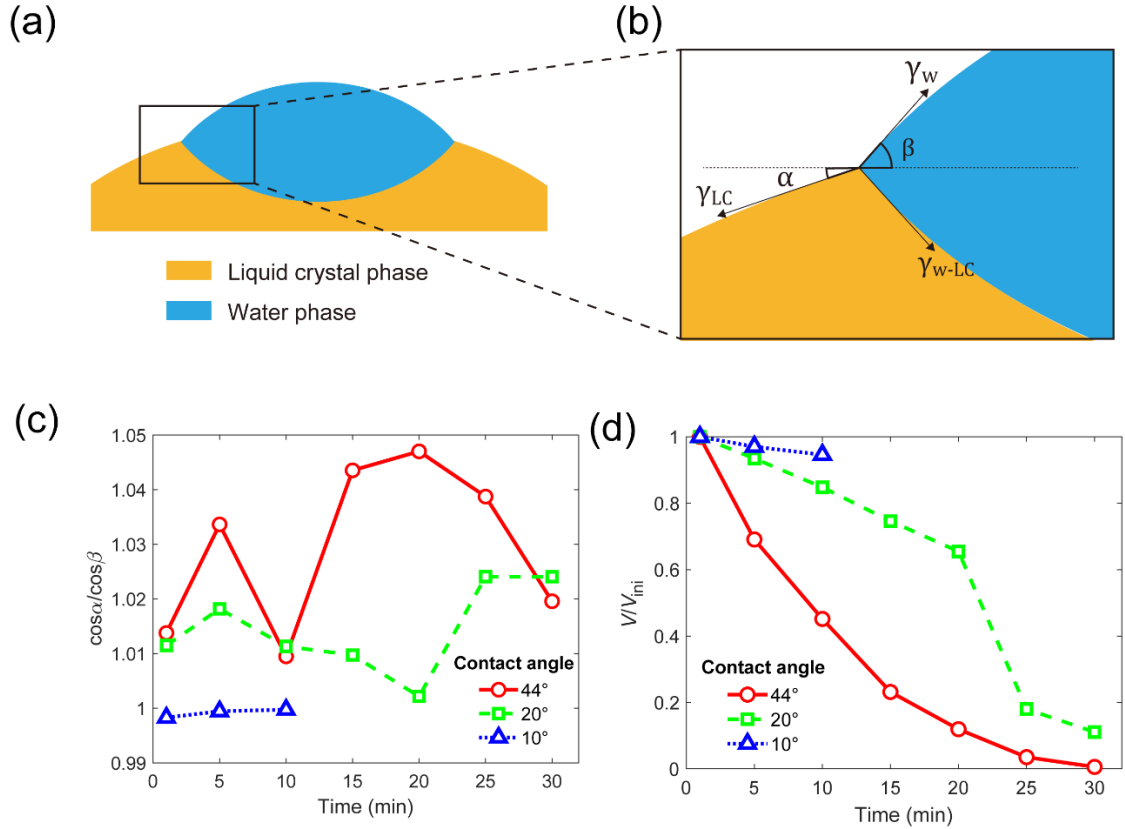

Figure 5. (a) Schematic representation of a cross-section of a water droplet on a 5CB droplet (b) Close-up of the contact line in (a). (c) Relative strength of surface tension between water and 5CB as a function of time for each wettability of glass slides. (d) Normalized volume of a water droplet as a function of time for each wettability of glass slides. $V$ and $V_{ini}$ are volume of a water droplet and volume of a water droplet at 1 min, respectively. In (c) and (d), data is missing when the contact angle is 10° because it was difficult to measure after 15 min.